\documentclass[manuscript]{emulateapj}

\slugcomment{to appear in the Astrophysical Journal}

\shorttitle{Recurrent Nova Model}
\shortauthors{Kato et al.}

\usepackage{apjfonts}

\begin{document}


\title{A SELF-CONSISTENT MODEL FOR A FULL CYCLE OF RECURRENT NOVAE -- WIND 
MASS LOSS RATE AND X-RAY LUMINOSITY}

\author{Mariko Kato} 
\affil{Department of Astronomy, Keio University, Hiyoshi, Yokohama
  223-8521, Japan;}
\email{mariko.kato@hc.st.keio.ac.jp}

\author{Hideyuki Saio}
\affil{Astronomical Institute, Graduate School of Science,
    Tohoku University, Sendai, 980-8578, Japan}

\and
\author{Izumi Hachisu}
\affil{Department of Earth Science and Astronomy, College of Arts and
Sciences, The University of Tokyo, 3-8-1 Komaba, Meguro-ku,
Tokyo 153-8902, Japan}



\begin{abstract} 
An unexpectedly slow evolution in the pre-optical-maximum phase 
was suggested in the very short recurrence period nova M31N 2008-12a. 
To obtain reasonable nova light curves we have improved our calculation 
method by consistently combining optically thick wind solutions of 
hydrogen-rich envelopes with white dwarf (WD) structures calculated 
by a Henyey-type evolution code.
The wind mass loss rate is properly determined with high accuracy.
We have calculated light curve models for 1.2 and 1.38 $M_\odot$ WDs 
with mass accretion rates corresponding to recurrence periods of
10 and 1 yr, respectively. 
The outburst lasts 590/29 days in which the pre-optical-maximum 
phase is 82/16 days, for 1.2/1.38 $M_\odot$, respectively.
Optically thick winds start at the end of X-ray flash 
and cease at the beginning of supersoft X-ray phase. 
We also present supersoft X-ray light curves 
including a prompt X-ray flash and later supersoft X-ray phase. 
\end{abstract}


\section{INTRODUCTION} \label{sec_introduction}

A nova is a thermonuclear runaway event on a mass-accreting white dwarf (WD)
\citep{nar80,ibe82,pri86,sio79,spa78}. 
Theoretical light curves of nova outbursts have been calculated mainly 
based on a steady-state approximation 
\citep[i.e., optically thick wind theory:][]{kat94h} 
with successful reproduction of 
characteristic properties, such as multiwavelength light curves including 
supersoft X-ray phase, of a number of classical novae 
\citep{hac06kb,hac07k,hac10,hac14ka,hac15k,hac16a,hac16b,hac08kc,kat09}. 
In spite of these successful results, this method is inapplicable 
when a steady-state is not a good approximation. 
Such failures occur for very fast evolutions in recurrent novae 
like U Sco and RS Oph and pre-maximum-optical phases of all kinds of novae. 
Thus, we need to take a different approach to follow 
the entire evolution of short period 
recurrent novae, which we aim in the present work. 

Multicycle nova outbursts have been calculated
with Henyey-type evolution codes.  These codes, however, 
meet numerical difficulties when the nova envelope expands to a giant size.
To continue numerical calculation beyond the extended stage, 
various authors have adopted various mass loss schemes and 
approximations 
\citep{pri95,kov98,den13,ida13,wol13a,wol13b,kat14shn,kat15sh,tan14}.
These works, however, paid little attention in reproducing reliable 
nova light curves, so there is still room for improvement.

Recently, X-ray flash of a nova, a brief X-ray brightening before the optical maximum, 
came into attention with realistic observational plans \citep{mor16,kat15sh,kat16xflash}.  
\citet{kat16xflash} showed that outbursts of recurrent novae are so weak that 
their pre-optical-maximum evolution could be very slow. 
They reported the nondetection of X-ray flux during 8 days 
prior to the 2015 outburst of M31N 2008-12a. 
They estimated that the X-ray flash had already occurred long ($9 - 16$ days)
before the optical maximum. 
Such unexpected slow evolution of the very early phase, in spite of the 
very fast optical decline, 
indicates the fact that we do not fully understand nova outbursts yet.

Recurrent novae of very short recurrence periods ($P_{\rm rec} < 10$ yr) 
were recently reported: 0.5 or 1 yr for M31N 2008-12a   
\citep{dar14,dar15,dar16errata, hen14, hen15,tan14,dar16}, 
5 yr for M31N 1963-09c \citep{wil15atel}, 
$\sim 6$ yr for LMCN 1968-12a \citep{dar16lmc68},  
and 10 yr for U Sco \citep[e.g.,][]{sch10}.
Recurrent novae are binaries harboring a massive white dwarf (WD). 
Short recurrence period novae occur in very massive WDs of
$M_{\rm WD}\gtrsim 1.2~ M_\odot$ with very high mass accretion rates of
$\dot M_{\rm acc}\gtrsim1.5 \times 10^{-7}M_\odot$~yr$^{-1}$
\citep{pri95,wol13a,wol13b,tan14,kat14shn}.
Such massive WDs are considered to be candidates for
Type Ia supernova (SN~Ia) progenitors
\citep{hknu99, hkn99, hac01kb, hkn10, han04, li97, kat12review}.
SNe Ia play very important roles in astrophysics as a standard candle
for measuring cosmological distances \citep{per99,rie98} 
and as main producers of iron
group elements in the chemical evolution of galaxies \citep{kob98}. 
However, their
immediate pre-explosion progenitors remain unclear 
\citep[e.g.,][]{mao14,pag14}.  
Thus, studies of very short recurrence period novae
are very important.

Our aim of this work is to establish a self-consistent numerical method for 
modeling a full cycle of nova outbursts. We combine 
the structure of optically thick winds with a model obtained by our 
Henyey-type code at each phase of a nova, 
which gives a mass loss rate and 
surface values of temperature and luminosity that are 
required to calculate multiwavelength light curves. 
In order to obtain successful fittings in our first attempt 
we have chosen a $1.2~M_\sun$ 
WD with a very high mass-accretion rate of 
$\dot M_{\rm acc} = 2.0 \times 10^{-7}~M_\odot$~yr$^{-1}$ because 
the nova explosion is weak and expansion is slow, so the 
envelope structure is close to that of steady-state envelope. 
We have also calculated wind structures fitted to the evolution models
of a $1.38~M_\sun$ WD with 1 yr recurrence period published in \citet{kat16xflash}.
Section \ref{sec_method} describes our numerical method,
and Section \ref{sec_m1.2} presents our results on the $1.2~M_\sun$ WD 
in detail. The $1.38~M_\sun$ WD model is briefly described in Section 
\ref{sec_m1.38}.
Discussion and conclusions follow in 
Sections \ref{sec_discussion} and \ref{section_conclusion}, respectively.

\section{NUMERICAL METHOD} \label{sec_method}

We calculated WD structures from the center of the WD up to 
the photosphere. 
We adopt the optically thick winds \citep[e.g.,][]{kat83,kat85,kat94h} 
as the main mechanism of mass loss during the nova outburst. 
The optically thick winds have been applied to a number of novae and 
successfully reproduced light curves and characteristic properties
\citep[e.g.,][]{hac06kb, hac07k, hac08kc,hac10,hac14ka,hac15k,
hac16a,hac16b,kat09}.  The interior structures of WDs are calculated 
with a Henyey code (Section \ref{sec_method_evolution}). 
When the optically thick winds occur the structure of  
mass-losing envelope is calculated as described 
in Section \ref{sec_method_wind}. 
We use the $UV$ plane for our fitting process of outer and inner structures 
(Section \ref{sec_method_UV}) in the way as described 
in Section \ref{sec_method_fitting}. 

\subsection{Time-dependent Calculation}
\label{sec_method_evolution}

Evolution models of mass-accreting WDs are
calculated by using the same Henyey-type code as in \citet{kat14shn,kat16sh} 
and \citet{hac16}.
Accretion energy outside the photosphere is not included. 
We have calculated multicycle nova outbursts until the characteristic 
properties of flashes become unchanged. 
The evolution is started with the equilibrium model for a given 
mass accretion rate
\citep{kat14shn}, in which the WD core is thermally in a steady state. 
Thus, the WD core temperature is not a free parameter. 

In a later phase of a shell flash, the photospheric radius (as well as
the luminosity) increases to the point where the outer boundary condition
of our evolution code fails because the density becomes too small
at $T\sim 2\times10^5$~K \citep[at the Fe opacity peak,][]{igl96}.
To avoid the numerical difficulty, we start mass loss, 
which hinders the growth of the radius,
when the photospheric radius $R_{\rm ph}^n$ at time-step $n$ 
exceeds a radius $R_0$. The mass loss rate for the next step
$\dot{M}_{\rm evol}^{n+1}$ is calculated as 
\begin{equation}
\dot{M}_{\rm evol}^{n+1} = \dot{M}_{\rm evol}^n\exp[10(\log R_{\rm ph}^n-\log R_{\rm ph}^{n-1})].
\label{eq:mdot}
\end{equation}
At the beginning of mass loss $\dot{M}_{\rm evol}^{n}$ is replaced with 
$\dot{M}_0$ ($<0$), which is an arbitrary parameter to give
an initial mass loss rate at $R_{\rm ph}=R_0$. 
This equation gives a rapid increase (decrease) in $|\dot{M}_{\rm evol}|$ 
when the photospheric radius is increasing (decreasing).
Equation\,(\ref{eq:mdot}) is used until $R_{\rm ph}$ becomes
smaller than a radius $R_1$, then $\dot{M}_{\rm evol}$ is switched to
the accretion rate given at the start of the calculation. 
In the $1.38~M_\sun$ model, we restarted the accretion immediately 
after the mass loss stopped.  This is because we adopt internal models 
of \citet{kat16xflash}. 
In the $1.2~M_\sun$ model, we restarted accretion when the outburst 
is almost finished, to compare the duration of the supersoft X-ray
phase with that obtained from the static sequence which does not include
mass-accretion (see Section \ref{sec_xraylightcurve}).

The values of the parameters, $\dot{M}_0$, $R_0$, and $R_1$ are chosen
case by case. The details of these parameters 
are explained in Section \ref{section_details}.

\subsection{Wind Mass Loss} \label{sec_method_wind}

Shortly after the onset of unstable nuclear burning, 
the envelope expands and optically thick winds are accelerated. 
The occurrence of winds is detected at the photosphere using the BC1 surface 
boundary condition in \citet{kat94h}. 
To obtain envelope structures with winds, 
we solve the equations of motion, mass continuity, radiative energy transfer 
by diffusion, and energy conservation assuming steady-state conditions and
spherical symmetry \citep{kat94h}.
The boundary conditions of the wind solution are set at 
the photosphere and critical point (=sonic point) 
\citep{kat94h}, which determines 
two values, i.e., the temperature and radius of the critical point,
($T_{\rm cr}$, $R_{\rm cr}$). 
We calculated the optically thick wind solutions 
down to a chosen inner boundary, i.e., where the temperature rises to 
a specified value (e.g. $T=T_{\rm f}$) and  
obtain the envelope mass of the wind ($M_{\rm env}^{\rm wind}$) between 
the photosphere and the point of $T=T_{\rm f}$. 
In short, for a chosen temperature ($T=T_{\rm f}$) a wind solution 
is characterized by $T_{\rm cr}$ and $R_{\rm cr}$. 
The other quantities are automatically calculated,
such as the mass loss rate, wind velocity, 
and local luminosity ($L_{r}$). In general $L_{r}$ is the  
sum of radiative and convective luminosities. In our calculation 
convection does not occur at the matching point and $L_{r}$ is 
always equal to the radiative luminosity.

\subsection{Fitting in the $UV$ Plane} \label{sec_method_UV}

In the fitting procedure we use the $UV$ plane
\citep{cha39,sch58,hhs62}, where $U$ and $V$ are the homology
invariants defined as
\begin{equation}
U \equiv {d \ln M_r \over {d \ln r}} = {4 \pi r^3 \rho \over M_r },
\label{equation_U}
\end{equation}
and
\begin{equation}
V \equiv -{d \ln P \over {d \ln r}} = {{G M_r \rho} \over {r P}}, 
\label{equation_V}
\end{equation}
where $M_r$ is the mass within the radius $r$.

In general, the fitting point of the wind solution ($U_{\rm w}$, $V_{\rm w}$)
moves downward 
in the $UV$ plane if we increase the wind mass loss rate 
(i.e., increase $T_{\rm cr}$ or $R_{\rm cr}$). 
On the other hand, the fitting point of Henyey-code solutions 
($U_{\rm f}$, $V_{\rm f}$) moves upward for a larger mass loss rate.
Thus, we find a unique matching point for the same wind mass loss rate and 
envelope mass in the $UV$ plane. 
Matching $U$ and $V$ at $T_{\bf f}$ guarantees the continuation of 
the mechanical structure at the fitting point so that the radius and density 
are automatically matched there (see Equations (\ref{equation_U}) and (\ref{equation_V})). 
In addition, as there is no energy source/sink around the fitting place 
the local luminosity $L_{r}$ is automatically fitted.

\subsection{Fitting Procedure} \label{sec_method_fitting}

In the wind phase we need several steps 
of iteration until we obtain a final result in which 
the wind mass loss is consistently taken into account. 
Our fitting procedure is as follows. 

\noindent
(1) First, we calculate nova outbursts with the Henyey code 
assuming a time-variable mass loss rate 
until the shell flashes reach a limit cycle. We use the last one cycle 
in the following procedure.   
\\  
(2) When the mass loss occurs on the Henyey code, we calculate   
optically thick wind solutions 
with a specified parameters, ($T_{\rm cr}$, $R_{\rm cr}$),
down to the fitting place $T=T_{\rm f}$. If the fitting condition 
($U_{\rm w}$, $V_{\rm w}$) = ($U_{\rm f}$, $V_{\rm f}$) is not satisfied, 
we change $T_{\rm cr}$ and $R_{\rm cr}$ until the condition is satisfied. 
We need typically 20 iterations done by hand. 
\\
(3) If the envelope mass $M_{\rm env}^{\rm wind}$ of the wind solution 
does not match to that of the interior solution we change the fitting 
place $T_{\rm f}$ and repeat the same process until the two envelope 
masses are sufficiently close to each other. 
A few to several iterations are needed for the process. 
\\
(4) We replaced the outer part ($T < T_{\rm f}$) of the evolution model 
with a well matched wind solution. 
Then, we obtain a continuous structure from the WD core up to 
the photosphere that consistently includes the optically thick wind 
at a specified time during the wind phase. 
The photospheric temperature, velocity, and wind mass loss rate etc. 
are uniquely determined.
\\
(5) We need to do such a procedure (2) - (4) for each timestep. 
However, our wind phase contains $\sim 10,000$ timesteps, which 
is far beyond the capacity of manual handlings. For this reason, we select 
about 30 epochs and apply the iteration procedure (2)-(4) by hand. 
Thus, we obtain stellar structure models for one cycle of nova outburst. 
\\
(6) The mass loss rates of the optically thick winds, 
which varies time to time, 
is generally different from those assumed in step (1). 
We repeated the process (1)-(5) assuming different mass loss rates 
$\dot M_0$ in Equation ({\ref{eq:mdot}) until the calculated 
wind mass loss rates well match to the
assumed values. It takes several iterations. 

Among the above process, the most difficult process is step (1). 
A Henyey-type code often meets numerical difficulties without assuming large 
mass loss rates when the envelope expands.  If we assume too large
mass loss rates, however, the procedure (1)-(6) does not converge. 

Note that the fitting point between the wind and the interior 
should be carefully chosen especially in the first trial. 
The fitting point should be deep enough to be insensitive to 
treatments in the surface region, especially when we assume inadequate 
mass loss rates in the trial process.
This point also should be shallow enough to ensure
the time-dependent term $\epsilon_{\rm g}$ 
\citep[gravitational energy release rate per unit mass: 
see equation (1) in][]{hac16} is negligible.

Our procedure (1)-(6) currently needs many iterations by hand, and 
thus, takes much human-time. 
For this reason, we have calculated only 1.2 and 1.38 $M_\sun$ models.  
We certainly need to improve our fitting procedure.

\section{1.2~$M_\odot$ Model} \label{sec_m1.2}

\begin{deluxetable}{llll}
\tabletypesize{\scriptsize}
\tablecaption{Summary of Recurrent Nova Models
\label{table_results}}
\tablewidth{0pt}
\tablehead{
\colhead{Subject\tablenotemark{a}}&
\colhead{units}&
\colhead{1.2 ~$M_\odot$ } &
\colhead{1.38 ~$M_\odot$ } 
}
\startdata
$\dot M_{\rm acc}$ & $10^{-7}~M_\odot$~yr$^{-1}$ & 2.0 & 1.6\\
$P_{\rm rec}$ & yr & 9.9   &  0.95\\
$\log T_{\rm core}$ & K  &  7.93 & 8.03\\
$M_{\rm acc}$ & $10^{-7}~M_\odot$  & 17  &1.4\\
$M_{\rm ig}$  & $10^{-7}~M_\odot$  & 26  & 2.0\\
$L_{\rm nuc}^{\rm max}$ & $10^6 L_\sun$  &2.2 &3.9    \\
$\log T_{\rm nuc}^{\rm max}$ & K  &8.11  & 8.23  \\
$t_{\rm nova}$\tablenotemark{b} & day & 590  &  29 \\ 
$t_{\rm wind}$ & day   &  130     &  19 \\
$t_{\rm X-flash}$\tablenotemark{c}& day  & 110     & 0.77   \\
$t_{\rm SSS}$\tablenotemark{c} & day  &  440    & 7.8    \\
$t_{\rm rise}$\tablenotemark{d} & day & 82  &  16 \\
$t_{\rm decay}$\tablenotemark{e} & day & 510  &  13
\enddata
\tablenotetext{a}{The WD mass and mass accretion rate are given 
parameters and recurrence period and other quantities are the 
results of calculation.
}
\tablenotetext{b}{duration of the flash with $L_{\rm ph} > 10^4 L_\sun$. }
\tablenotetext{c}{duration of the X-ray flash and SSS phase: 
$L_{\rm X} (0.3-1.0$ keV) $>10^3 L_\sun$ for 1.2$~M_\sun$, 
but $>10^4 L_\sun$ for $1.38~M_\sun$.}
\tablenotetext{d}{ the time elapsed between the time of 
the peak nuclear luminosity and the time of the peak mass loss rate. }
\tablenotetext{e}{optical decay time calculated as 
$t_{\rm decay}=t_{\rm nova}-t_{\rm rise}$.}
\end{deluxetable}

We have calculated two models of 1.2 and 1.38~$M_\odot$. 
The 1.38~$M_\odot$ model are almost 
the same as those partly published in \citet{kat16xflash}.  
Thus, we first describe the 1.2~$M_\odot$ model 
and then summarize for the 1.38~$M_\odot$ results focusing on the difference. 
For the mass loss parameters of the $1.2~M_\odot$ model discussed in Section \ref{sec_method_evolution} 
we adopt $\log R_0/R_\sun=-1.25$ 
(corresponding to the case that we start mass loss at $\log T_{\rm ph}$ (K)= 5.54) 
and $\log R_1 /R_\sun = -1.36$ 
(corresponding to the case that we stop mass loss at $\log T_{\rm ph}$ (K)=5.60). 
 The resultant mass loss rates are shown later in 
Figure \ref{light.m12}.

Table \ref{table_results} summarizes the results of our calculations. 
The WD mass and mass-accretion rate are given parameters. 
The others are the results of calculation: the recurrence period $P_{\rm rec}$, 
temperature at the WD center $T_{\rm core}$, 
mass accreted after the previous outburst 
until the beginning of the current outburst 
(=mass transferred from the companion)  
$M_{\rm acc}$,    
ignition mass $M_{\rm ig}$ which is the envelope mass at the epoch 
of maximum nuclear luminosity 
$L_{\rm nuc}^{\rm max}$, maximum temperature at the epoch of 
maximum nuclear luminosity $T_{\rm nuc}^{\rm max}$, 
nova outburst duration $t_{\rm nova}$ for $L_{\rm ph} > 10^4~L_\sun$, 
duration of the wind phase $t_{\rm wind}$, 
durations of the X-ray flash $t_{\rm X-flash}$, and SSS phase $t_{\rm SSS}$, rising time $t_{\rm rise}$ which is the time elapsed between the 
time of the peak nuclear luminosity and the time of 
the peak mass loss rate, 
decaying time $t_{\rm decay}=t_{\rm nova}-t_{\rm rise}$. 
Here, the durations of the X-ray flash and SSS phase are defined as 
the time span when $L_{\rm X} > 10^4~L_\sun$ 
for 1.38 $M_\sun$ followed by \citet{kat16xflash}, 
but $L_{\rm X}> 10^3~L_\sun$ for 1.2 $M_\sun$ because of 
faint X-ray luminosity, where $L_{\rm X}$ means the 
luminosity in a band of $0.3 - 1.0$ keV. 
Note that the outburst duration $t_{\rm nova}$ is different from 
the summation of the three phases, $t_{\rm X-flash}+t_{\rm wind}+t_{\rm SSS}$, 
because the definition is different.  

\begin{figure}
\epsscale{1.15}
\plotone{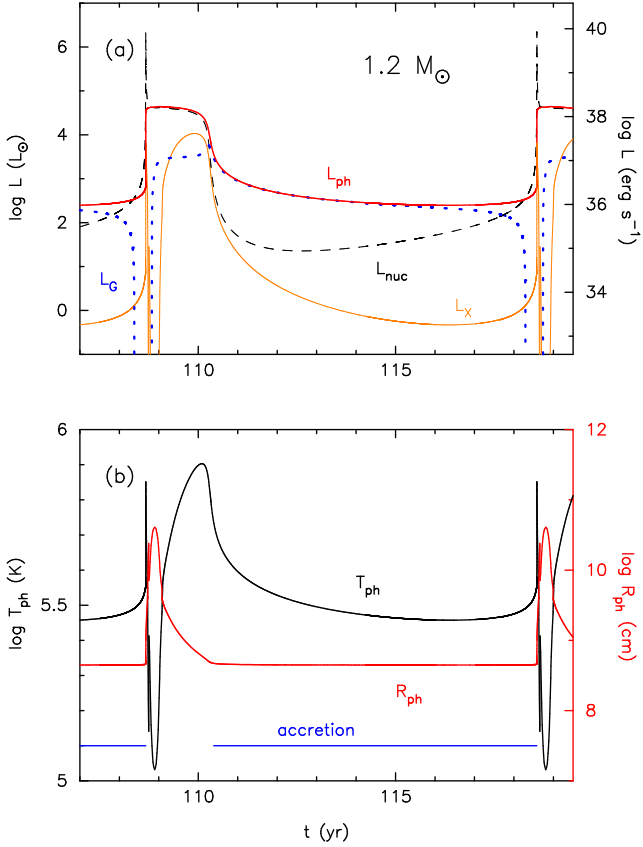}
\caption{Temporal change in one cycle of a nova outburst on 
a $1.2~M_\sun$ WD with a mass accretion rate of 
$2 \times 10^{-7}~M_\sun$~yr$^{-1}$.  
(a) The photospheric luminosity $L_{\rm ph}$ (red solid line), 
integrated flux of nuclear 
burning $L_{\rm nuc}$ (black dashed line), integrated 
gravitational energy release rate $L_{\rm G}$ (blue dotted line), 
and supersoft (0.3 - 1.0 keV) X-ray flux (solid orange line). 
(b) The photospheric temperature $T_{\rm ph}$ (black solid line) 
and radius $R_{\rm ph}$ (red solid line). The period of accretion 
 is indicated by 
the horizontal blue line.  
\label{light.Long.m12}}
\end{figure}

Figure \ref{light.Long.m12} shows the last cycle 
of our 1.2 $M_\odot$ WD model.  
We stopped the mass accretion 
($\dot M_{\rm acc} = 2.0 \times 10^{-7}~M_\odot$~yr$^{-1}$) 
when the luminosity increases to $\log L_{\rm ph}/L_\sun=4.32$ 
($\log T_{\rm ph}$ (K)= 5.85), 
and resumed mass accretion after the flash when the luminosity 
decreases to $\log L_{\rm ph}/L_\sun=4.0$ ($\log T_{\rm ph}$ (K)$= 5.83$).
The period of mass accretion is depicted in 
Figure \ref{light.Long.m12}(b). 
The recurrence period is obtained to be 9.9 yr.

Figure \ref{light.Long.m12}(a) shows 
the photospheric luminosity, $L_{\rm ph}$,
nuclear burning energy release rate integrated from the center
of the WD to the surface, $L_{\rm nuc}$, 
and total gravitational energy release rate, $L_{\rm G}$, 
defined as the integration of $\epsilon_{\rm g}$ from the center
of the WD to the photosphere \citep[equation (3) in ][]{kat16xflash}. 
The supersoft X-ray (0.3 $-$ 1.0 keV) luminosity $L_{\rm X}$ 
is calculated assuming the blackbody emission of photospheric 
temperature $T_{\rm ph}$.  
The lower panel (b) shows the change of the photospheric
temperature $T_{\rm ph}$ and radius $R_{\rm ph}$.

\begin{figure}
\epsscale{1.15}
\plotone{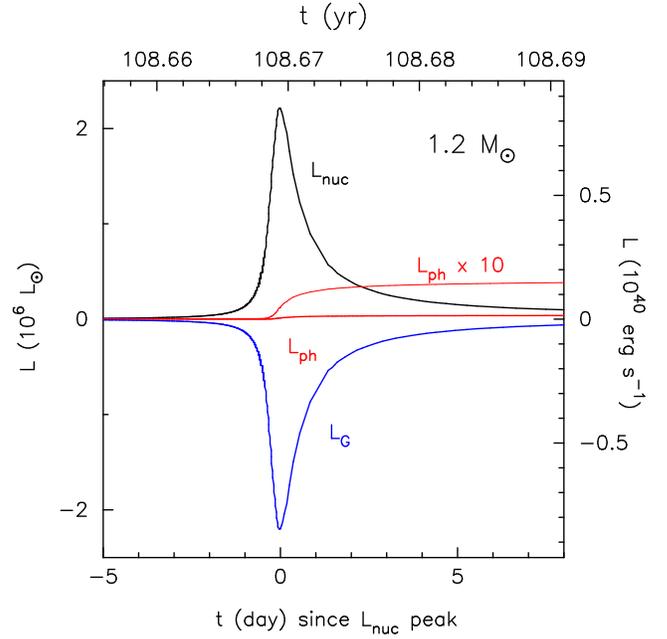}
\caption{Closeup view of a very early phase of the shell flash in 
Figure \ref{light.Long.m12}: the nuclear burning luminosity (black line), 
$L_{\rm nuc}$, photospheric luminosity (thick red line),  $L_{\rm ph}$, 
and gravitational energy release rate (blue line), $L_{\rm G}$. 
Most of the nuclear luminosity is absorbed
in the burning shell as expressed by large negative values of $L_{\rm G}$.
As a result, the photospheric
luminosity, $L_{\rm ph}$, is much smaller than $L_{\rm nuc}$.
The thin red line denotes 10 times the photospheric luminosity,
$L_{\rm ph} \times 10$.
\label{Lnuc.m12}}
\end{figure}

A closeup view in a very early phase is shown in Figure \ref{Lnuc.m12}.
The nuclear energy release rate reaches $L_{\rm nuc}= 2.2\times 10^{6} 
L_\odot$ at maximum.  
This energy is mostly absorbed in the lower part of the burning region as 
shown by a large negative value of $L_{\rm G} (< 0)$.  
Thus, only a very small part of 
$L_{\rm nuc}$ is transported outward up to the photosphere.  
As a result, the photospheric luminosity $L_{\rm ph}$ 
does not exceed the Eddington luminosity. 
These  properties are the same as those reported in the $1.38~M_\sun$
model \citep[Figure 2 in][]{kat16xflash}. 
Comparing the two models in Table \ref{table_results}, 
we find that $L_{\rm nuc}^{\rm max}$ of the 1.2~$M_\sun$  
model is almost a half of the $1.38~M_\sun$ model, i.e., 
the outburst is very weak.

The absorbed energy ($L_{\rm G} < 0$) is released in the later phase of 
the outburst. As shown in Figure \ref{light.Long.m12}(a), 
$L_{\rm G}~(>0)$ reaches about 10 \% of $L_{\rm ph}$ during 
the outburst. 
After the shell flash ends, $L_{\rm nuc}$ quickly decreases, 
but $L_{\rm ph}$ slowly decreases because the mass-accretion resumes
and $L_{\rm G}$ contributes to $L_{\rm ph}$. 
During the quiescent phase $L_{\rm ph}\approx L_{\rm G}$ is almost constant 
($\log L_{\rm ph}/L_\sun \gtrsim 2.4$) 
as shown in Figure \ref{light.Long.m12}(a).

\subsection{H-R Diagram}\label{sec_hr}

\begin{figure}
\epsscale{1.15}
\plotone{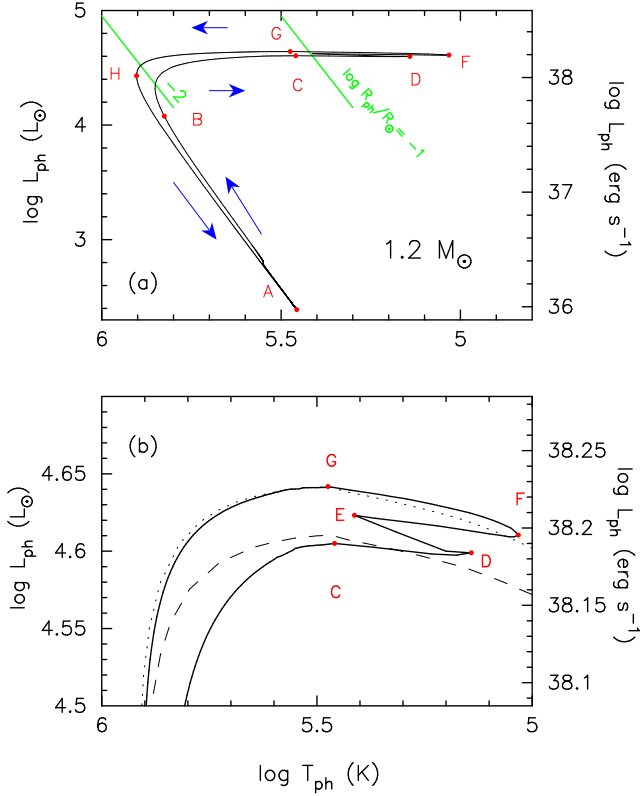}
\caption{Evolution track of the model in Figure \ref{light.Long.m12}.
Filled circles on the track correspond to the characteristic 
stages. A: minimum photospheric luminosity,  $L_{\rm ph}=L_{\rm ph}^{\rm min}$.  
B: maximum nuclear luminosity $L_{\rm nuc}=L_{\rm nuc}^{\rm max}$. 
C: optically thick winds start. 
D: first local maximum expansion. 
E: the uppermost hydrogen-rich envelope of $X=0.7$ is blown off. 
F: maximum expansion of the photosphere.
G: optically thick winds cease. 
H: maximum photospheric temperature $T_{\rm ph}=T_{\rm ph}^{\rm max}$. 
(a) The blue arrows indicate the direction of evolution.
The green solid lines indicate the lines of $\log R_{\rm ph}/R_\sun=-1$ and $-2$ (left).
(b) Closeup view. 
The thin surface layer of $X=0.7$ is blown off 
in the optically thick winds through C-D-E. 
Afterward the envelope composition is almost 
uniform at $X \sim 0.6$.  
The dotted and dashed lines indicate steady-state sequences with the chemical composition of   
$X=0.6,~Y=0.38$ and $Z=0.02$, and  $X=0.7,~Y=0.28$ and $Z=0.02$, 
respectively (see Section \ref{sec_m1.2}).
\label{hr.m12}}
\end{figure}

Figure \ref{hr.m12} shows the track in the H-R diagram for the same 
period in Figure \ref{light.Long.m12}. 
During the quiescent phase the accreting WD model stays around point A.
After the unstable nuclear burning sets in, the bolometric
luminosity quickly increases whereas the photospheric radius stays
almost constant (from point A to B).  
The nuclear energy generation rate $L_{\rm nuc}$
reaches its maximum at point B.  As the envelope expands, the photospheric
temperature decreases and optically thick winds start to blow at point C. 
The winds continue until it reaches point G. 
At point F the photospheric temperature attains a minimum.
After that the photospheric radius  decreases and the temperature 
$T_{\rm ph}$ increases with time.  The envelope mass decreases owing 
to wind mass loss and hydrogen burning. 

\begin{figure}
\epsscale{1.15}
\plotone{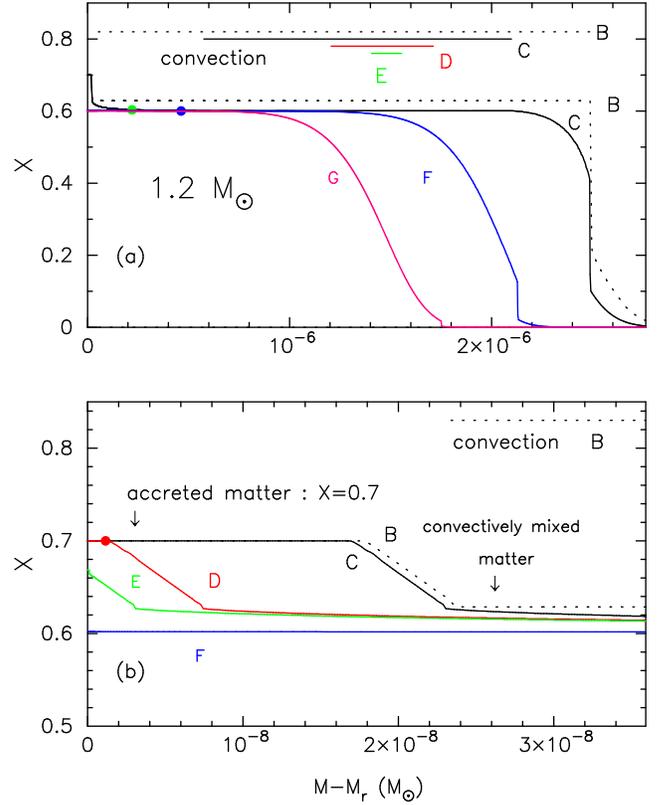}
\caption{
Runs of the hydrogen mass fraction, $X$, in the WD envelopes at selected stages. 
(a) $X$ distribution in the entire hydrogen-rich envelope. 
(b) Closeup view of the surface region of panel (a). 
In the both panels, the left bound ($M-M_r=0$) corresponds to
the photosphere. The accreted matter has a hydrogen content of 
$X=0.7$ initially. 
The envelope mass decreases owing to wind mass loss and nuclear burning.
Each stage is indicated beside the line. 
The outermost $X=0.7$ layer is blown off just after stage E.
The  convective region is indicated by the horizontal lines in the upper 
part of panel (a), which becomes narrower and disappears just after stage E.
The matching point of the outer wind solution with the inner structure is 
indicated by filled circles.
\label{X.m12}}
\end{figure}

Figure \ref{hr.m12}(b) shows a closeup view of the same track. 
There is a zigzag part (D-E-F) arising from the change 
of chemical composition in the surface layers. 
Just after the thermonuclear runaway sets in, convection develops 
throughout the envelope up close to the photosphere. 
In the uppermost layer where the convection never reaches,  
the chemical composition remains to be the same as that 
of the accreted matter, i.e., 
$X=0.7,~Y=0.28$ and $Z=0.02$. In the region where the convection reached 
the hydrogen mass-fraction decreased to $X \sim 0.6$. 
Figure \ref{X.m12} shows distributions of hydrogen mass-fraction in the 
envelope at selected stages, 
in which all the layer with $X=0.7$ have been blown off in the wind a little after
stage E. 
The gradient of $X$ results in the opacity gradient 
(for larger $X$, opacity is larger owing to electron scattering) 
that causes additional acceleration of the winds. 
Thus, the winds blow relatively stronger until the region with $X=0.7$ 
is completely blown 
off. As the chemical composition distribution approaches uniform 
winds become weak and the photospheric temperature rises with time. 
As the shell flash goes on the envelope expands and the 
photospheric temperature decreases again. 
Thus, the track in the H-R diagram has a zigzag excursion 
before it reaches point F in Figure \ref{hr.m12}(b).  
The envelope mass continuously decreases through these stages. 
Note that the distribution of hydrogen abundance $X$ 
in Figure \ref{X.m12} is that of the evolution model. 
In wind solution we assumed X-profile mimic to the corresponding evolution model.

The dotted and dashed lines in Figure \ref{hr.m12}(b) 
are the steady-state/static 
sequences that represent the nova decay phase  
\citep{kat99,kat94h}. The dashed line is for a $1.2~M_\sun$ WD 
with the uniform chemical composition of 
$X=0.7,~Y=0.28$ and $Z=0.02$ while the  
dotted is for $X=0.6,~Y=0.38$ and $Z=0.02$. 
The WD radius, which is needed in obtaining steady-state models, i.e., 
the radius at the bottom of the hydrogen-rich envelope, 
is taken from our evolution calculation as $\log R/R_\sun = -2.22$. 

In the decay phase (stages F, G, and later) the evolution path is very 
close to that of the upper steady-state sequence (dotted line). 
In this phase, the envelope composition is almost uniform at 
$X=0.6$, $Y=0.38$, and $Z=0.02$, and the envelope structure is close to 
that of the steady-state (stage F to G) and of 
hydrostatic balance (after stage G) as well as thermal equilibrium. 
Thus, the nova evolves along with the steady-state/static sequence. 

In the rising phase, on the other hand, the envelope does not yet reach 
thermal balance but approaches the balance with time. 
The envelope still has an $X=0.7$ layer on its top, so 
the evolution path is close to the lower dashed line.

\subsection{Envelope structure}

\begin{figure}
\epsscale{1.15}
\plotone{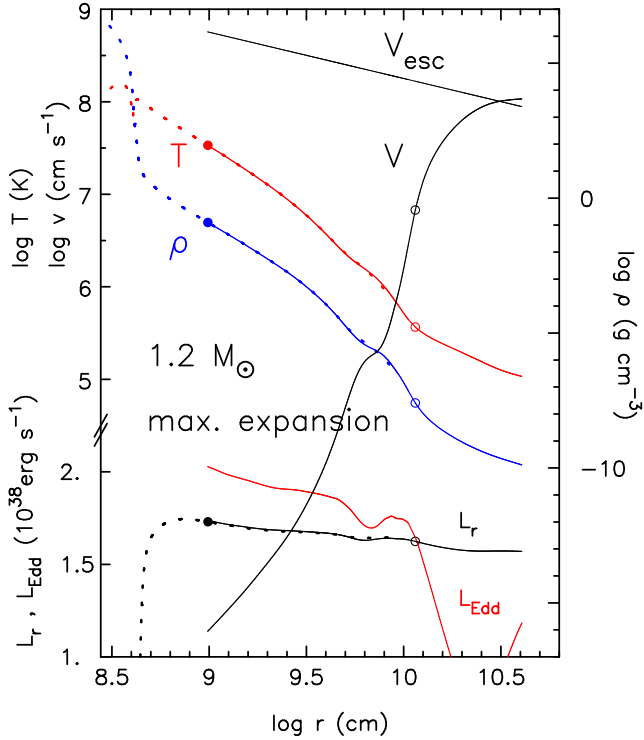}
\caption{Internal structure at the maximum expansion of the photosphere  
(point F in Figure \ref{hr.m12}).
We plot, top to bottom, the escape velocity $v_{\rm esc}$
(thin solid black line), wind velocity $v$ (solid black line),
temperature $T$ (red line), density $\rho$ (blue line), 
local luminosity $L_r$ (black line),
and local Eddington luminosity $L_{\rm Edd}$ (red line).
The solid lines indicate the optically thick wind solution, whereas
the dotted lines show the inner structure obtained by the Henyey code.
The local Eddington luminosity, shown only from the outside
of the fitting point (filled circle), becomes smaller
than the radiative luminosity for $\log r$ (cm) $> 10.1$.
The interior solution is plotted beyond the fitting point 
to demonstrate that the inner and outer solutions are 
very close to each other. 
The open circles denote the critical point of the wind solution \citep{kat94h}.
Convective region has disappeared before this stage. 
\label{struc.m12.max}}
\end{figure}

Figure \ref{struc.m12.max} shows 
the envelope structures of the wind solution (solid lines) and Henyey code solution 
(dotted lines) at the maximum expansion of the photosphere (epoch F).
The photosphere reaches $\log R_{\rm ph}$ (cm)=10.61 ($0.58~R_\sun$). 
This figure also shows the distribution of the local 
luminosity $L_{\rm r}$ 
and the local Eddington luminosity defined by
\begin{equation}
L_{\rm Edd} = {4\pi cG{M_{\rm WD}} \over\kappa},
\label{equation_Edd}
\end{equation}
where $\kappa$ is the opacity. We adopt OPAL opacity \citep{igl96}. 
The local
Eddington luminosity has a small dip corresponding to
a small peak in the OPAL opacity owing to ionized O and Ne 
($\log T$ (K) $\sim 6.2$--6.3), and a large decrease corresponding to
a large Fe peak at $\log T$ (K) $\sim 5.2$ 
\citep[see Figure 6 of][]{kat16xflash}.
The wind is accelerated where the opacity increases outward.
The wind velocity barely exceeds the escape velocity
$v_{\rm esc}=\sqrt {2 G M_{\rm WD}/r}$.

The dotted lines show the interior structure of the model calculated 
with the Henyey code 
in step (1) in Section \ref{sec_method_fitting}.
The connecting point with the wind solution is 
depicted by filled circles. 
To demonstrate the smooth fitting, we plot the model structure 
beyond the connecting point.  
It shows a good agreement with the steady state solution 
not only on the connecting point but also in the wide outside region. 

\begin{figure}
\epsscale{1.15}
\plotone{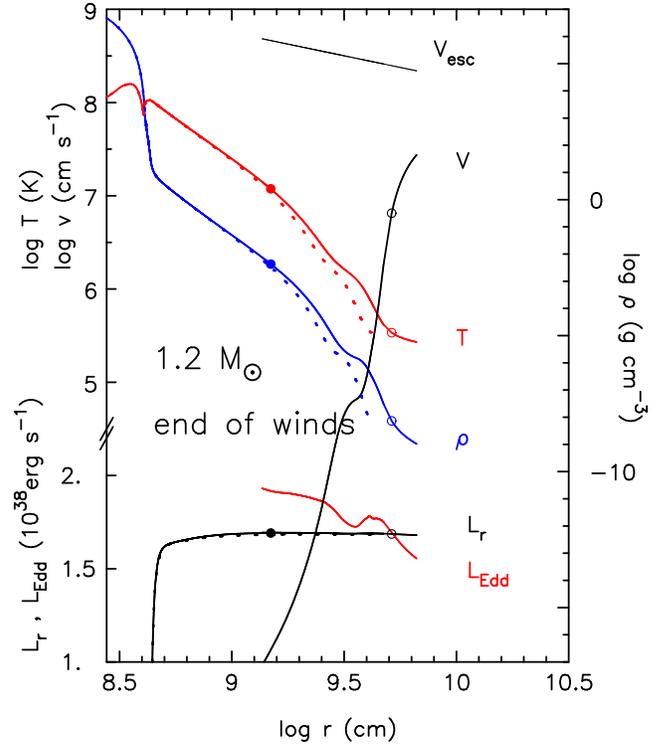}
\caption{
Same as Figure \ref{struc.m12.max}, but just before (solid) and 
at (dotted) stage G.  The wind becomes weaker compared with that at 
stage F in Figure \ref{struc.m12.max}. 
The local super-Eddington region becomes much narrower. 
At stage G, when the optically thick wind stops (dotted line), 
there is no local super-Eddington region. No convection occurs.
\label{struc.m12.lsmax2}}
\end{figure}

Figure \ref{struc.m12.lsmax2} demonstrates how the optically thick 
winds 'quietly' cease at stage G. 
In the decay phase of the nova outburst, the photospheric temperature 
increases with time and the dip of $L_{\rm Edd}$ becomes shallower. 
The solid line depicts the envelope structure 
just before stage G. The wind is still accelerated, but 
the acceleration region, where $L_r > L_{\rm Edd}$, is getting narrower. 
This region disappears at stage G (dotted line). 
The resemblance of the two structures demonstrates 
that the optically thick winds gradually weaken  
and stop quietly. 
This property was already pointed out in the 
steady-state sequences \citep{kat94h}  
that the optically 
thick winds quietly ceases when the photospheric temperature rises 
to beyond the Fe peak.

\subsection{X-ray Light Curve and Wind Mass Loss}
\label{sec_xraylightcurve}

\begin{figure}
\epsscale{1.15}
\plotone{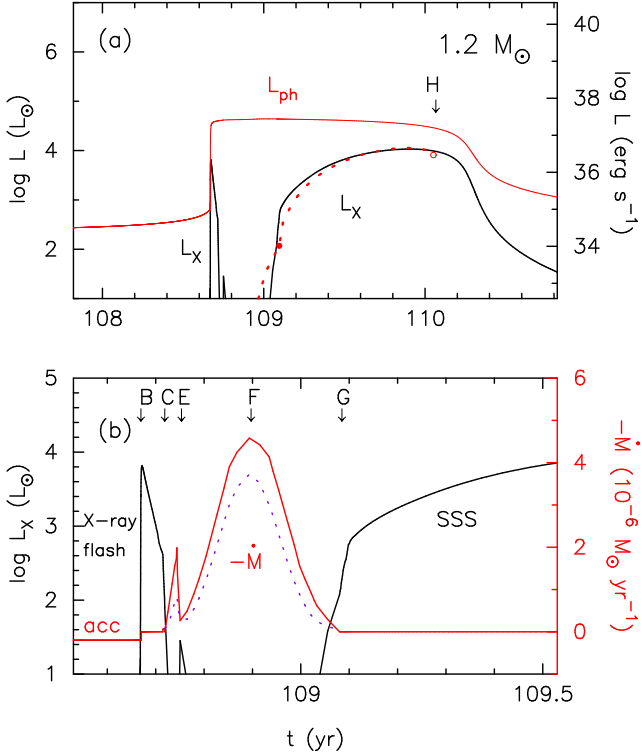}
\caption{(a) Temporal change of supersoft (0.3 - 1.0 keV) X-ray luminosity $L_{\rm X}$ (black line) 
and photospheric luminosity $L_{\rm ph}$ (solid red line) of the $1.2~M_\sun$ WD model.  
The dotted red line denotes the X-ray light curve of steady-state sequence, in which the wind stops 
at the small filled circle and hydrogen nuclear burning stops at the small open circle 
(see section \ref{sec_xraylightcurve}).
Epoch of stage H is indicated. 
(b) Closeup view of the first half part of the upper panel 
with the mass loss rate of optically thick wind $\dot M_{\rm wind}$ (red line). 
Dotted purple line is the assumed mass loss rate $\dot M_{\rm evol}$ in process (1) 
in Section \ref{sec_method_fitting}.  
Stages B, C, E, F, and G are indicated. 
\label{light.m12} }
\end{figure}

In Figure \ref{light.m12} the upper panel (a) shows the bolometric 
and supersoft X-ray 
luminosities during the outburst, while the lower panel (b) shows 
its closeup view with wind mass loss rates. 
The overall one-cycle view is already shown in Figure \ref{light.Long.m12}. 
The first bright X-ray phase is the X-ray flash that occurs just after 
the onset of thermonuclear runaway \citep{kat16xflash}. 
The tiny X-ray peak shortly after the X-ray flash 
corresponds to the small leftward excursion toward stage E in 
the H-R diagram (Figure \ref{hr.m12}). 
The supersoft X-ray phase begins at stage G and ends shortly after stage H.

The panel (b) also shows the temporal change of the wind mass loss rate 
which is closely related to the change in the X-ray light curve. 
The X-ray flash ends when the envelope expands and the 
optically thick winds starts at stage C. As the photospheric radius 
increases (photospheric temperature decreases) toward stage F, 
the wind mass loss rate increases with time. 
After stage F the mass loss rate decreases as the 
star evolves back from stage F to G in the H-R diagram (Figure \ref{hr.m12}). 
The wind stopped at stage G when the SSS phase begins. 
In this way the X-ray turn on/off is closely related to the occurrence of the optically thick winds.  
The mass loss rate increases (decreases) as the photospheric temperature 
decreases (increases) which will be shown later (Figure \ref{dmdtT} in 
Section \ref{sec_masslossrate}).

A sequence of static and steady-state solutions is also shown in  
Figure \ref{light.m12}(a) for comparison.  
The dotted red line depicts the X-ray light curve 
calculated from the model shown by the dotted line 
in Figure \ref{hr.m12}. The rightmost point 
corresponds to the epoch of extinction of hydrogen burning.  
This X-ray light curve shows a good agreement with the evolutional one 
but the termination point comes earlier. 
This difference is explained as follows. The evolution time of the equilibrium  
sequence is calculated from the mass decreasing rate owing to nuclear burning. 
On the other hand, in the time-dependent calculation, the evolution time is 
influenced by the gravitational energy release at the bottom of the envelope, 
i.e., the evolution time is determined by 
the mass decreasing rate owing to nuclear burning and 
additional energy source of gravitational energy release rate $L_{\rm G}$. 
This additional energy source, amounts about $10 \%$ of 
the photospheric luminosity (Figure \ref{light.Long.m12}),
lengthens the SSS lifetime.
Considering this effect we can conclude that the steady-state sequence 
is consistent with the time-dependent calculation.

\section{1.38 $M_\sun$ Model}\label{sec_m1.38}

Our multicycle shell flash calculation of a 1.38 $M_\odot$ 
WD with $\dot M_{\rm acc} = 1.6 \times 10^{-7}~M_\odot$~yr$^{-1}$ was 
published \citep{kat16xflash}, but it included only the 
early phase of the X-ray flash. 
Using the method of calculation discussed in Section \ref{sec_method_fitting} 
we improved the fitting in the wind mass loss phase. 
We adopt $\log R_0/R_\sun= -1.00$ and $\log R_1/R_\sun= -1.15$.
The recurrence period is 0.95 yr. 
The characteristic properties of this model 
are summarized in Table \ref{table_results}.  
Here, we report our results focusing on 
the difference from the 1.2 $M_\odot$ case.

\begin{figure}
\epsscale{1.15}
\plotone{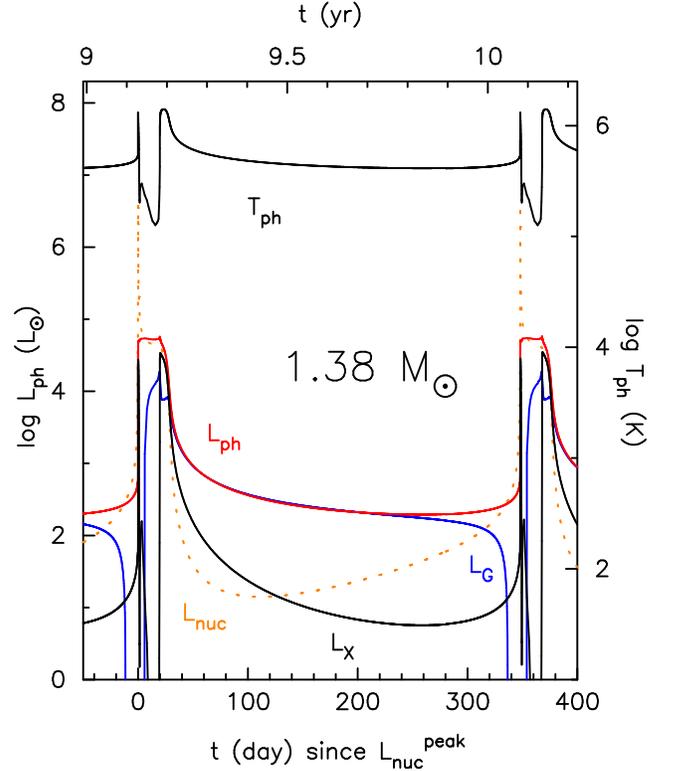}
\caption{Same as those in Figure \ref{light.Long.m12}, but 
for the $1.38~M_\sun$ model with a mass accretion rate of 
$1.6 \times 10^{-7}~M_\odot$ yr$^{-1}$.
\label{LT.m138}}
\end{figure}

Figure \ref{LT.m138} shows the last one cycle of our 
calculation. 
The outburst duration ($L_{\rm ph} > 10^4~L_\sun$) is 29 days. 
In the interpulse phase, the gravitational energy release rate $L_{\rm G}$ 
dominates the photospheric luminosity $L_{\rm ph}$. 
The energy budget in the early phase, corresponding to Figure \ref{Lnuc.m12},  
is already published in \citet{kat16xflash}. These characteristic properties 
are same as those in the 1.2 $M_\odot$ WD model.

\begin{figure}
\epsscale{1.15}
\plotone{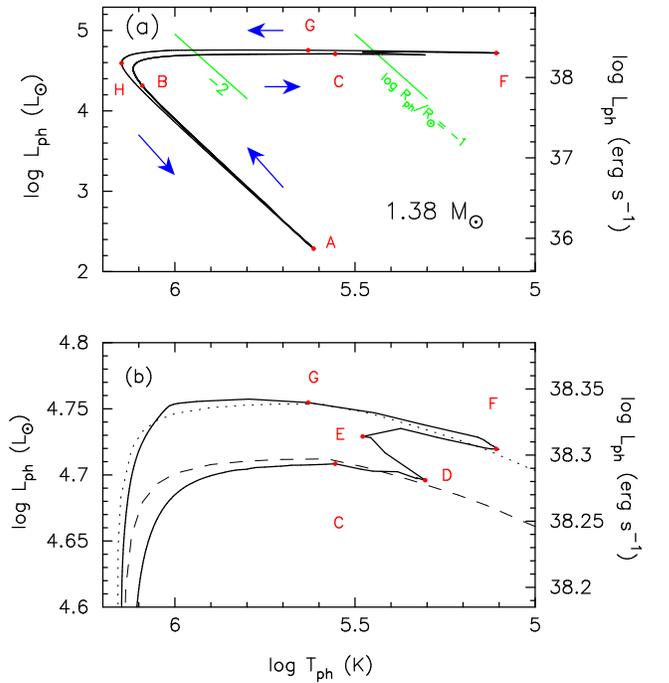}
\caption{
Same as those in Figure \ref{hr.m12}, but for the $1.38~M_\sun$ model.
In panel (b), the dotted and dashed lines indicate the sequences
of steady-state/static solutions of the 
chemical composition of $X=0.55,~Y=0.43$, and $Z=0.02$, and 
$X=0.7,~Y=0.28$, and $Z=0.02$. 
In these sequences, the WD radius is assumed to be 
$\log R/R_\sun=-2.57$ (taken from our evolution calculation).
See the main text for more details.
\label{hr.m138}}
\end{figure}

Figure \ref{hr.m138}(a) shows one cycle of nova outbursts in the H-R diagram. 
A closeup view is shown in panel (b). 
Similarly to the 1.2 $M_\odot$ model (Figure \ref{hr.m12}), 
stages C-D-E correspond to 
the period that the uppermost layer with $X=0.7$, of mass 
$1.6 \times 10^{-9}~M_\sun$, are blown off in the wind. 
After that, the chemical composition in the envelope is almost 
uniform at $X=0.58$, $Y=0.4$ and $Z=0.02$, and the envelope 
evolves along with the line of steady-state/static sequence.

\begin{figure}
\epsscale{1.15}
\plotone{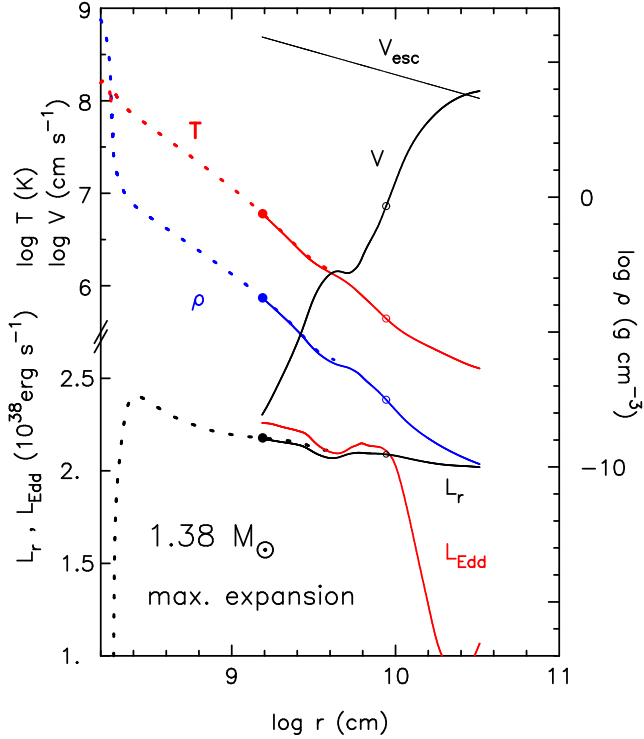}
\caption{Same as those in Figure \ref{struc.m12.max}, but for the 
$1.38~M_\sun$ model. Convective region has disappeared before this stage. 
\label{struc.m138.maxexp}}
\end{figure}

Figure \ref{struc.m138.maxexp} shows the internal structure
at stage F. The velocity quickly rises (around $\log r$ (cm) $\sim$10.0) 
where the local Eddington 
luminosity quickly decreases corresponding to the Fe opacity peak. 
The envelope mass at this stage is as small as $1.1 \times 10^{-7}~M_\sun$
which is not enough to expand to a giant size, so the photospheric radius 
is as small as $\log R_{\rm ph}$~(cm) =10.52 ($ 0.48~R_\sun$), 
and the surface temperature is rather high at $\log T_{\rm ph}$ (K)=5.10.

To demonstrate a good matching between the interior and wind solutions, 
the interior structure outside the fitting point is also plotted; 
this outer part was replaced with the wind solution in step (4) in Section \ref{sec_method_fitting}. 
The dotted line is very close to the steady-state wind solution, indicating 
that the interior structure is smoothly connected 
to the wind solution. 

\begin{figure}
\epsscale{1.15}
\plotone{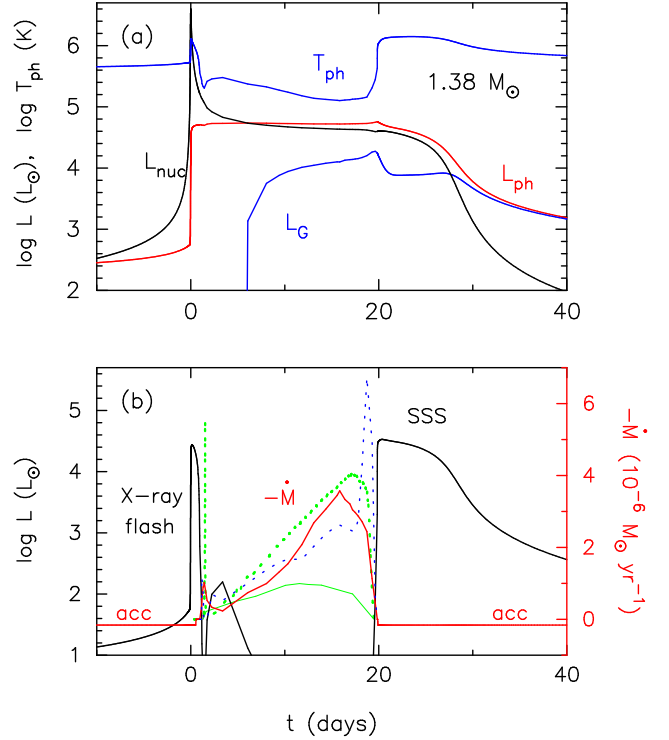}
\caption{(a) The nuclear burning luminosity, $L_{\rm nuc}$ (black line),
photospheric luminosity, $L_{\rm ph}$ (red line), 
rate of energy release owing to gravitational contraction, $L_{\rm G}$ (lower blue line), 
and photospheric temperature, $T_{\rm ph}$ (upper blue line) in the outburst phase 
of the $1.38~M_\sun$ model.
(b) The X-ray luminosity for the 0.3 - 1.0 keV band, $L_{\rm X}$ (black line), and  
wind mass loss rate $\dot M_{\rm wind}$ (red line).
The mass loss rate $\dot M_{\rm evol}$ initially assumed in step (1) in Section \ref{sec_method_fitting} is 
shown by the dotted blue line. 
Another trial iteration is added: $\dot M_{\rm evol}$ (dotted green line)
and $\dot M_{\rm wind}$ (solid green line). For these two lines the time is normalized 
to fit the wind phase of the red line model. 
See Section \ref{section_details} for more details. 
\label{light.m138}}
\end{figure}

Figure \ref{light.m138}(a) shows the temporal 
changes in $L_{\rm ph}$, $L_{\rm nuc}$, and $L_{\rm G}$ as well as the 
change of $T_{\rm ph}$.  The supersoft X-ray light curve shown in 
panel (b) is calculated 
from $L_{\rm ph}$ and $T_{\rm ph}$. 
The X-ray flash (when $L_{\rm X} > 10^4 ~L_\sun$) 
lasts 0.77 days and the SSS phase 7.8 days. 
Between them there is a faint X-ray peak corresponding 
to stage E.  

This model may be compared with the observational properties of the 
1 yr recurrence period nova
M31N 2008-12a. \citet{kat15sh} estimated the WD mass of M31N 2008-12a
to be $1.38~M_\sun$ from the SSS duration that lasts 
about 8 days \citep{hen14,hen15,dar16}.
In the present paper, we do not consider the optical light curves because 
the relation between the optical brightness and wind mass loss rate 
in recurrent novae is 
uncertain, which remains to be a future work. 
If we assume that the optical magnitude peaks at stage F, i.e., when the wind 
mass loss rate is maximum, the X-ray turn-on time is 4 days after the 
optical maximum, which 
is roughly consistent with M31N 2008-12a \citep[$5.9 \pm 0.5$ days:][]{hen15}, 
considering the insufficient fitting quality of our $1.38~M_\sun$ model. 
Detailed comparison with the observation 
is beyond the scope of the present paper.

\section{DISCUSSION}\label{sec_discussion}

\subsection{Wind Mass Loss Rate} \label{sec_masslossrate}

\begin{figure}
\epsscale{1.15}
\plotone{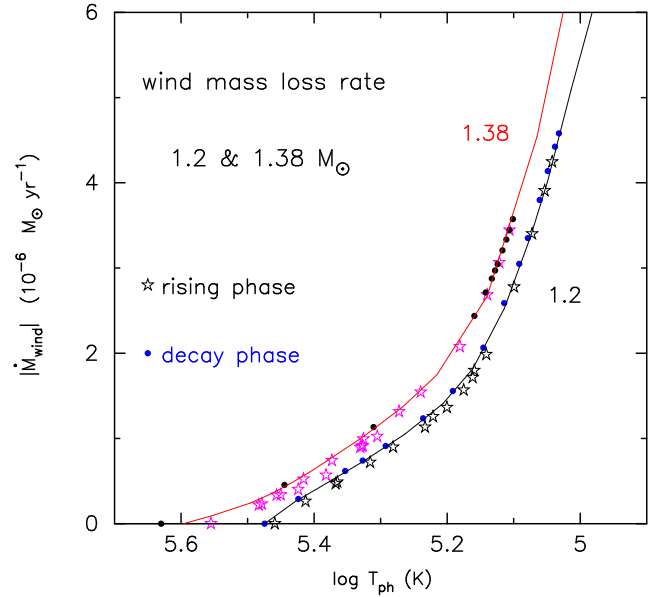}
\caption{
Mass loss rates of optically thick winds against the photospheric
temperature for the 1.2 and 1.38~$M_\odot$ models.
The open star symbols indicate from stage C to F 
in Figures \ref{hr.m12} and \ref{hr.m138}, 
whereas filled circles indicate at and after stage F. 
The thin lines depict the relation obtained from the steady-state sequences 
for 1.2 and 1.38~$M_\odot$ with the chemical composition of
$X=0.6$, $Y=0.38$, and $Z=0.02$ (lower black line) and 
$X=0.55$, $Y=0.43$, and $Z=0.02$ (upper red line), respectively. 
\label{dmdtT}}
\end{figure}

Figure \ref{dmdtT} shows the mass loss rate of the optically thick winds 
calculated in the fitting process (Section \ref{sec_method_fitting}) 
against the photospheric temperature $T_{\rm ph}$. 
The mass loss rate increases as $T_{\rm ph}$ 
decreases. In the extended envelopes 
the acceleration region (where the opacity increases outward) 
locates deep below the photosphere 
where the density is large (see Figures \ref{struc.m12.max} and \ref{struc.m12.lsmax2}),  
which results in a large mass loss rate. 
The mass loss rate also weakly depends on the WD mass. 
The $1.38~M_\sun$ model shows a slightly 
higher mass loss rates because the surface gravity is stronger in 
more massive WDs. 

This figure also shows the two $\dot M_{\rm wind}$ - $T_{\rm ph}$ 
relations (solid lines) obtained from the steady-state sequences of 
1.2 and 1.38 $M_\sun$, corresponding 
to the upper sequence in each H-R diagram (Figure \ref{hr.m12}(b) for 
1.2 $M_\sun$ and Figure \ref{hr.m138}(b) for 1.38 $M_\sun$). The dependence on 
the chemical composition is very weak, so the 
solar composition sequences are omitted. 
The mass loss rates obtained from our time-dependent calculations 
agree very well with 
those of steady-state sequences throughout the wind phase, including 
the zigzag path, C-D-E-F-G.  

This agreement of the $\dot M_{\rm wind}$ - $T_{\rm ph}$ relations 
is reasonable because the interior structure 
of our time-dependent calculation shows a good agreement
with the steady state solution. 
The envelope is accelerated where the opacity rapidly increases outward
at $T \sim 2\times 10^5$ K, so that 
similar envelope structures result in similar photospheric temperatures 
and wind mass-loss rates.

\subsection{Details of Numerical Calculation} \label{section_details}

In Section \ref{sec_method_evolution} we described the mass loss rate 
$\dot M_{\rm evol}$ which is initially assumed.  
After several time-consuming iterations, we found that better results 
are obtained if we choose $R_0$ to be close to those where the optically thick wind 
stops in the steady-state sequence \citep{kat94h}, i.e., when the photospheric
temperature is close to that where $\dot M_{\rm wind}=0$ in Figure \ref{dmdtT}.  

Figure \ref{light.m12}(b) shows $\dot M_{\rm evol}$ 
(dotted purple line) and resultant wind 
mass-loss rate (solid red line) in the $1.2~M_\sun$ model. 
The resultant wind mass loss rates are somewhat larger than the assumed value 
($ 19~\%$ at the time of maximum wind mass loss) but 
traces well the global change including early small peak 
in the episode of surface composition change (see Section \ref{sec_hr}). 
This agreement is satisfactorily well, considering numerical difficulties in step (1). 

Figure \ref{light.m138}(b) shows $1.38~M_\sun$ models for two sets of 
$\dot M_{\rm evol}$ (dotted lines) and corresponding resultant 
(solid lines) mass loss rates $\dot M_{\rm wind}$. 
Blue dotted line show $\dot M_{\rm evol}$ obtained with 
$\log R_0 /R_\sun = -1.0$ and $\log R_1 /R_\sun = -1.15$,  which results in the wind 
mass loss rate shown by the solid red line. 
The $25 \%$ matching is obtained at the maximum mass loss rate (of the red line)
but there is a short period of a large discrepancy of factor 2 
just before the winds stop. 

Dotted green line shows $\dot M_{\rm evol}$ obtained adopting 
$\log R_0/R_\sun =-1.5$ and $\log R_1 /R_\sun =-1.3$. In this case 
$\dot M_{\rm evol}$ has no large 
enhancement in the final wind phase, but a sharp narrow peak in the very early stage. 
The resultant wind mass loss rates, $\dot M_{\rm wind}$, denoted by the solid green line, 
however, are much small in the later phase. 
To summarize, with a smaller $\dot M_{\rm evol}$ 
(dotted blue line) we get 
a slightly larger $\dot M_{\rm wind}$ except the final peak, 
and with a larger $\dot M_{\rm evol}$ 
(dotted green line), we get a much smaller $\dot M_{\rm wind}$. 
We suppose that a well converged model should have wind mass loss rates between 
them, which may be closer to the red line model than the solid green line. 
Therefore, we adopted the red line model.

\subsection{Comparison with other works}

In the present work we calculated nova evolution models with very fine 
mass zonings and short timesteps. As a results, we are able to follow 
the temporal change of the wind mass loss, including the gradual increase 
and decrease, as in Figures \ref{light.m12} and 
\ref{light.m138}. We also showed how the upper envelope with $X=0.7$ 
is blown off (Figure \ref{X.m12} and zigzag tracks in Figures \ref{hr.m12} and \ref{hr.m138}). 

Many time-dependent calculations on novae have been presented so far, but 
none of them showed a zigzag behavior in the H-R diagram.
For example, \citet{pri95} calculated nova multicycle outbursts
for various WD masses, 
of which light curves are recently published by \citet{hil14}.  
Their light curves show no zigzag behavior like ours. 
The difference might arise from the difference in zonings of the 
envelope, because if the mass zoning did not resolve well the 
superficial $X=0.7$ region, 
the model would be unable to follow the temporal change of the 
chemical composition of the superficial layers. 
This reasoning may be supported by the description ``convection had already reached 
the surface of the star'' \citep[Section 2.1][]{pri95}. 
In addition, in their nova outbursts, the photospheric temperature suddenly 
jumps to the minimum value, and stay there, and jumps back. This corresponds to, 
in our Figure \ref{hr.m12}, that the star jumps from stage C to F 
instantaneously and jumps back to stage G. 
Such a behavior could occur if large timesteps were adopted, skipping  
the gradual increase/decrease periods of the mass loss rates.

\citet{kov98} presented an algorithm to place
an optically thick wind solution on top
of a hydrostatic stellar configuration. 
The calculation shows no indication of the zigzag shape. 
This is because the adopted small number of mass zones (less than 50)
is not sufficient to treat the very thin $X=0.7$ layer.  
Despite the detailed description of the algorithm, no 
information on the connecting point is given. 
In the 1.0 $M_\sun$ model, the wind begins at $\log T_{\rm eff}$ (K)=5.36 
and ceases at  $\log T_{\rm eff}$ (K)=5.5, which is roughly consistent with our 
model with the OPAL opacity. At the occurrence of the wind the mass loss rate 
suddenly jumps from zero to the maximum value followed by a gradual decrease. 
The gradual decrease is consistent with our calculation. 

\citet{ida13} adopted Kovetz' (1998) mass loss scheme with  
sufficient number of mass zones ($ > 8000$), however, the paper discusses 
interior structures in detail with little description of the surface region.

In the previous paper \citep{kat15sh} we presented a multi-wavelength 
light curve model of $1.38~M_\odot$ WD 
for the 1 yr recurrence period nova M31N 2008-12a. We adopted a similar 
method of numerical calculation to the present work, but connected 
the interior structure at a much deeper place with the wind solutions.
One of the reason why we chose such fitting point was that our 
numerical calculation with the Henyey code did not converge
unless large mass loss rates were assumed, and the fitting place 
was chosen for the structure to be 
almost independent of the assumed mass loss rate. 
However, in such a deep envelope with a larger mass loss rate, the 
gravitational energy release rate per unit mass
$\epsilon_{\rm g}$ is not negligible, so the fitting of steady-state 
solutions with the interior yielded mediocre models. 
The resultant mass loss rates were about 3 times larger 
than the present work. Thus, the wind phase was 
shorter by a factor of $\sim 3$. 
This roughly explains the difference between the previous estimate of 
a duration of 5 days and our present estimate of 19 days. 

In the present work we have improved the numerical technique and 
adopt much thinner atmosphere and mass zoning, and connect to the interior 
structure at a place where $\epsilon_{\rm g}$ is negligible. 
As a result, matching of inner and outer solutions are sufficiently 
well as shown in Figures \ref{struc.m12.max} 
and \ref{struc.m138.maxexp}. Our $1.2~M_\sun$ model 
shows very good convergence of iteration process 
(Section \ref{sec_method_fitting}) throughout the wind phase 
that support our fitting procedure. 
Our $1.38~M_\sun$ model shows a better convergence except 
a short period just before the  
SSS phase. This model must be improved further in the future.

\section{CONCLUSIONS}\label{section_conclusion}

Our main results are summarized as follows.

\noindent
1. We present a new calculation procedure to follow nova outbursts in which 
the optically thick wind is consistently introduced. 
We have found that steady-state wind solutions smoothly match with 
the interior calculated with the Henyey-type code.

\noindent
2. We calculated one cycle nova outburst on 1.2 and 1.38~$M_\sun$ WDs with 
the mass-accretion rates corresponding to the recurrence periods of
$P_{\rm rec}=10$ and 1 yr, respectively.  
The wind mass loss rate is accurately obtained in the new calculation procedure. 
The X-ray flash ends when the optically thick winds begin. 
The SSS phase starts when the winds cease.
At the maximum expansion of the photosphere, its temperature is as high as $10^5$ K 
in the both cases.

\noindent
3. We obtained a zigzag track in the H-R diagram corresponding 
to the epoch that the very surface zone of $X=0.7$ is blown off in the wind, uncovering the region mixed by convection during the runaway, 
which therefore has a lower H fraction correspondingly different opacity.

\noindent
4. Our $1.38~M_\sun$ model is roughly consistent 
with the timescale of the 1 yr recurrence period nova M31N 2008-12a. 

\noindent
5. The relation between the wind mass loss rate and photospheric temperature 
in our evolution model agrees well with those of steady-state/static sequence. 
The envelope structure at each stage are also very similar to those
of steady-state/static sequence in the wide region 
except the deep interior where the gravitational energy release rate per
unit mass $\epsilon_{\rm g}$ is not negligible.

\acknowledgments
MK and IH express our gratitude to T. Iijima and Astronomical Observatory of Padova 
(Asiago) for warm hospitality during which we initiated the present work. We also thank the anonymous referee for useful comments that 
improved the manuscript. 
This research has been supported in part by Grants-in-Aid for
Scientific Research (15K05026, 16K05289)
of the Japan Society for the Promotion of Science.


%

\clearpage


















\end{document}